\documentclass[style=bcnpostgrad,entry,preprint]{proceedings}
\renewcommand{\conferencetitle}{Barcelona Postgrad Encounters on Fundamental Physics}
\usepackage[utf8]{inputenc}
\usepackage{amsthm}
\usepackage{amsmath}
\usepackage{amsfonts}
\usepackage{amssymb}
\usepackage{mathrsfs}
\usepackage{tabularx}

\newtheorem{Theorem}{Theorem}
\newtheorem{Definition}{Definition}

\begin{document}

% check your name, e-mail address and institution
\author{Andrzej Borowiec, \underline{Aneta Wojnar}}
\email{borow@ift.uni.wroc.pl, aneta@ift.uni.wroc.pl}
\affiliation{Instytut Fizyki Teoretycznej, Uniwersytet Wroclawski,\\ pl. M. Borna 9, 50-204 Wroclaw, Poland}

\title{Geometry of almost-product Lorentzian manifolds and relativistic observer} % required

\abstract{The notion of relativistic observer is confronted with Naveira's classification of (pseudo-)Riemannian
almost-product structures on spacetime manifolds. Some physical properties and their geometrical counterparts are shortly discussed.} % required

\acknowledgements{A.W. would like to thank the Organizers for warm hospitality during the Conference.\newline
The calculations have been partially performed in Maxima.} % optional

\maketitle

\section{Introduction}

In Einstein's General Relativity, a gravitational interaction is represented by a metric with Lorentzian signature \((-,+,+,+)\)
living on a (curved) four-dimensional spacetime manifold and satisfying Einstein's field equations. An observer is an  independent
notion and, according to a nowadays point of view, can be identified with an arrow of time. More precisely, the observer is
determined by a timelike normalized (local) vector field on spacetime.
We can also think of it as the collection of its integral
curves, considered as world lines (also known as the congruence of world lines of point observers) of some continuous material object
(e.g. relativistic fluid). From a mathematical perspective, it provides a one-dimensional (timelike) foliation.
It appears that a pair, the metric and the vector field, determines a differential-geometric structure which is called an
almost-product structure. From a physical perspective, a relativistic observer is tautologically defined as a field of his own
four-velocities. Having chosen an observer, one can define relativistic observables, i.e. relative measurable quantities. They include the
relative (three-)velocity of another observer or test particles (see e.g. \cite{bini},\cite{oziewicz}, \cite{ungar}), as well as Noether conserved currents in diffeomorphism
covariant field theories \cite{mauro}. The well-known splitting of the electromagnetic field into measurable electric and magnetic
components is also relative to the observer. In the more traditional approach to General Relativity, the measurable quantities are
related to coordinates. In fact, given a coordinate system, one can associate to it a (local) observer, indicated by a time variable.
However the notion adopted here is more general, coordinate-free and can be globalized.\\
In the presented note we provide the correspondence between Naveira's classes of a pseudo-Riemannian manifold \cite{naveira}
implemented by the observer and its physical characteristics as introduced in \cite{ehlers}.\\
The paper is organized as follows. In section 2 we introduce the notation and basic notions. In section 3 we shortly recall
Gil-Medrano's theorem \cite{gil}, which provides a differential geometric interpretation for Naveira's classes. The advantages of
the almost-product structures in physics are discussed in section 4 (see also \cite{coll} in this context). They extend the possible
characteristics for a given observer on a Lorentzian manifold. Finally, we provide a few illustrative examples in section 5.

\section{Preliminaries and definitions}

 Let \(M\) and \(TM\) denote respectively an \(n\)-dimensional smooth manifold and its tangent bundle. % possibly with a pseudo-Riemannian metric and the corresponding Levi-Civita connection \(\nabla\).
A \(k\)-dimensional (\(k<n\)) tangent distribution (\(k\)-distribution in short) is a map \(D\) which associates
a \(k\)-dimensional subspace \(D_p\subset T_pM\) to the point \(p\in M\):
\begin{equation}
 D:\;p\;\rightarrow\;D_p\subset T_pM.
\end{equation}
$D$ can be also considered as a subbundle of $TM$.
Locally, one can say that a \(k\)-distribution is generated by a set of \(k\) linearly independent vector
fields iff in every point  \(p\) their values span the
\(k\)-dimensional subspace \(D_p\), i.e.\\ \(D_p=\textrm{span}\{X_1(p),\ldots, X_k(p)\}\).
In this case we shall write \(X_i\in \Gamma(D)\), where $\Gamma(D)$ stands for a submodule of cross sections of the subbundle $D\subset TM$.\\
An embedded submanifold \(N\subset M\) is called an integral manifold of the distribution \(D\) if \(T_pN=D_p\) in every
point \(p\in N\).
We say that \(D\) is involutive if, for each pair of local vector fields \((X,Y)\) belonging to \(D\), their Lie bracket \([X,Y]\) is
also a vector field from \(D\).\\
The distribution \(D\) is completely integrable if for each point \(p\in M\) there exists an integral manifold \(N\) of the
distribution \(D\) passing through \(p\) such that the dimension of \(N\) is equal to the dimension of \(D\).
It turns out that every involutive distribution is completely integrable (local Frobenius theorem). Every smooth \(1\)-dimensional
distribution is integrable.\\
The integrability of a distribution is closely related to the notion of foliation. We have the following (global) Frobenius theorem:
\begin{Theorem}
 Let \(D\) be an involutive \(k\)-dimensional tangent distribution on a smooth manifold \(M\). The collection of all maximal
connected integral manifolds of \(D\) forms a foliation of \(M\).
\end{Theorem}
The proof of the theorem and the precise definition of a foliation can be found in \cite{lee}. Roughly speaking, a foliation is a
collection of submanifolds \(N_{i}\) such that each submanifold proceeds smoothly into another one. They do not cross each other.
Particularly, a class of globally hyperbolic spacetimes $M=T\times\Sigma$, where $T$ is an open interval in the real
line $\mathbb{R}$ and $\Sigma$ is  a three-manifold, serve as a typical example of global foliation \cite{hawking}.\\
Let us recall \cite{gray, yano} that an almost-product structure on \(M\) is determined by a field of endomorphisms of \(TM\), i.e. a \((1,1)\) tensor
field \(P\) on \(M\), such that \(P^2=I\) (\(I= \)Identity). In this case, at any point \(p\in M\), one can consider two subspaces
of \(T_pM\) corresponding respectively to two eigenvalues $\pm 1$ of \(P\).
%\(1\) and \(-1\), \(D_p\) and \(D^\perp_p\) respectively which
It defines two complementary  distributions on \(M\), i.e. $TM=D^+\oplus D^- $.
Moreover, if \(M\) is equipped with a (pseudo~-)Riemannian metric \(g\) such that
\begin{equation}\label{comp}
 g(PX,PY)=g(X,Y);\;\;X,Y\in\Gamma(TM),
\end{equation}
then both distributions are mutually orthogonal. In this case, \(P\) is called a (pseudo~-) Riemannian almost-product structure.
It is to be noticed that some modified gravity models admit almost-product structures as solutions \cite{borow}.

\section{Geometric characterization of distributions on (pseudo-)Riemannian manifolds}

Let \(D\) be a distribution on \((M,g)\) and \(D^{\perp}\) the distribution orthogonal to \(D\). At every point \(p\in M\), we
have then \(T_pM=D_p\oplus D_p^{\perp}\)
\footnote{The case of null distributions is more complicated and should be discussed separately, see e.g. \cite{bejancu}.}.
Thus we can uniquely define a \((1,1)\) tensor field \(P\) such that \(P^2=I,\; P_{|D}=1,\; P_{|D^{\perp}}=-1\). It is clear
that $P$ becomes automatically a (pseudo~-)Riemaniann almost-product structure. One has (see \cite{gil}):

\begin{Definition}
The distribution \(D\)  is called geodesic, minimal or umbilical if and only if \(D\)  has property \(D_{1}\), \(D_{2}\) or \(D_{3}\)
respectively, where:
\begin{itemize}
\item \(D_{1}\Longleftrightarrow\;(\nabla_{A}P)A=0\),

\item \(D_{2}\Longleftrightarrow\;\alpha(X)=0\),

\item \(D_{3}\Longleftrightarrow\; g((\nabla_{A}P)B,X)+g((\nabla_{B}P)A,X)=\frac{2}{k}g(A,B)\alpha(X)\),
\end{itemize}
where \(X\in \Gamma(D)^{\perp};\; A,B\in \Gamma(D).\)
Here \(\{e_{a}\}_{a=1}^{k}\;\)\((k=\mathrm{dim}D)\) is a local orthonormal frame of \(D\)
and \(\alpha(X)={\displaystyle \sum_{a=1}^{k}g((\nabla_{e_{a}}P)e_{a},X)}\).
\end{Definition}
 It implies that a distribution has the property \(D_1\) if and only if it has the properties \(D_2\) and \(D_3\). Their meanings
in the case of integrability are explained below.
\begin{Theorem} (O. Gil-Medrano)
A foliation \(D\) is called totally geodesic, minimal or totally umbilical if and only if \(D\) has the property \(F_1\), \(F_2\) or \(F_3\)
respectively, where
\begin{equation}
 F_i\,\iff\,F+D_i,\;\;i=1,2,3
\end{equation}
and
\begin{equation}
 F\,\iff\,(\nabla_A P)B=(\nabla_B P)A\;\;\forall\,A,B\in\Gamma(D).
\end{equation}
\end{Theorem}
The proof of this theorem can be found in \cite{gil}.
It is easy to see that the property \(F\) is equivalent to Frobenius' theorem, i.e. a distribution \(D\) with this property
is a maximal foliation. The theorem says that, in principle, one deals with three special types of foliations:\\
($F_1$) Totally geodesic foliation:  it means that every geodesic of an arbitrary integral submanifold $N$ (the leaf of foliation),
if considered together with the induced metric (the first fundamental form), is at the same time geodesic of the total manifold \(M\).
Moreover, it is equivalent to the statement that the second fundamental form of \(N\)
(i.e. extrinsic curvature) vanishes. In other words, the extrinsic curvature measures the failure of a geodesic of the
manifold \(N\) to be a geodesic of \(M\).\\
($F_2$) Minimal foliation: If there is a surface with the smallest possible value of the area bounded by a certain curve, that surface
is called a minimal surface. The condition for a distribution to be a minimal distribution is that the trace of the second fundamental
form vanishes. The trace of the extrinsic curvature is also called mean curvature, that is, the average of the principal curvatures.
Examples of  minimal surfaces in \(\mathbb{R}^3\) are the catenoid and the helicoid.\\
($F_3$) Umbilical foliation: We recall that an umbilical manifold  is a manifold for which all points are umbilical points.
Umbilical points, in turn, are locally spherical: every tangent vector at such point is a principal direction and all principal
curvatures are equal \cite{spivak4}. For example, a sphere is an umbilical manifold. In the case of integral submanifolds, the second
fundamental form has to be proportional to the induced metric.

\section{Almost-product structure related to a spacetime observer}
 In the present section we are going to apply the formalism presented above to the special case of a relativistic observer on a
spacetime manifold.  These new tools will be used at the end of the section for a final classification.\\
From now on $(M, g)$ denotes  a four-dimensional manifold (spacetime) equipped with Lorentzian signature metric $g_{\alpha\beta}$.
An observer is represented by a timelike vector field $u^\alpha$ which, according to our sign convention \((-,+,+,+)\), is normalized to
\begin{equation}\label{norm}
 u^\alpha u_\alpha=-1.
\end{equation}
Strictly speaking, the normalization condition (\ref{norm}) prevents the existence of critical points and one can deal with
a one-dimensional (timelike) distribution instead. Such a distribution is always integrable and provides a foliation with world-lines
as leaves. Each leaf can then be parameterized by arc length (proper time), making $u^\alpha$ a four-velocity field. This implies
that the only nontrivial question one can ask about a one-dimensional distribution is weather it is geodesic or not (see below the tables).\\
Because of this, one should concentrate on its orthogonal (transverse) completion $D$. This is a spacelike three-dimensional
distribution with Euclidean signature. These two distributions provide a $3+1$ (orthogonal) decomposition of the tangent
bundle $TM=D\oplus D^\perp$, with the one-dimensional timelike distribution denoted as $D^\perp$. It is easy to find out that
the corresponding three-dimensional projection tensor has the form:
\begin{equation}\label{proj}
 h^\alpha_{\,\beta}=\delta^\alpha_{\,\beta} + u^\alpha u_\beta,
\end{equation}
which, due to (\ref{norm}), implies $h^\alpha_{\,\rho} h^\rho_{\,\beta}= h^\alpha_{\,\beta}$. We would like to stress that in
what follows we shall always use the original metric $g_{\alpha \beta}$ for lowering and rising indices. Thus covariant and
contravariant components of  tensors can be used exchangeably. For example, the second-rank symmetric tensor
\begin{equation}\label{ind_metric}
 h_{\alpha\,\beta}=g_{\alpha\,\beta} + u_\alpha u_\beta,
\end{equation}
plays the role of induced Euclidean metric on the distribution $D$. When $D$ is integrable, then (\ref{ind_metric}) is the
first fundamental form (i.e. induced metric) on each leaf. The  corresponding foliation by spacelike hypersurfaces has the
physical meaning of clock synchronization and divides the spacetime into  equal-time pieces identified as three-dimensional spaces.
One should mention that the integrability of $D$ is always required in the case of $3+1$ splitting which is necessary for the
Hamiltonian formalism of General Relativity (see e.g. \cite{eric}).\\
More generally, to any tensor $A^{\alpha\cdots}_{\beta\cdots}$ living in the spacetime one can assign its projected three-dimensional
counterpart
\begin{equation}\label{proj2}
\tilde A^{\alpha\cdots}_{\beta\cdots}=h^\alpha_{\,\mu} h^\nu_{\,\beta}\cdots A^{\mu\cdots}_{\nu\cdots}
\end{equation}
According to widely spread ideas (see e.g. \cite{hawking, ehlers, poisson}), only projected three-dimensional  tensors are good candidates for measurable relativistic
observables. Obviously, such quantities are relative, i.e. observer dependent. For example, for an anti-symmetric
covariant two-tensor $F_{\alpha\beta}=-F_{\beta\alpha}$ (two-form), which under the closeness condition ($dF=0$) can be interpreted
as an electromagnetic field, one gets
\begin{equation}\label{em}
F_{\alpha\beta}= H_{\alpha\beta}+u_\alpha E_\beta-u_\beta E_\alpha\,,
\end{equation}
where $H_{\alpha\beta}=\tilde F_{\alpha\beta}=h^\mu_{\,\alpha}h^\nu_{\,\beta}F_{\mu\nu}$ and $E_\alpha=u^\mu h^\nu_\alpha F_{\mu\nu}$ are measurable electric and magnetic
components.

%  We will try to determine a geometrical object what is an almost-product structure with notions which have physical interpretations. It
%turns out that a metric, which represents a gravitational interaction, and an observer described by a normalized timelike vector field
% can be use to construct the almost-product structure which will satisfy necessary conditions.

%There are a few equivalent ways of thinking about the observer on Lorentzian manifold. We are working with the metric of signature \((-,+,+,+)\) so
%the normalisation of our observer has a form:

%We are using the metric tensor \(g_{\mu\nu}\) representing gravitational field to raise and lower indicies.
%According to a nowadays point of view the observer can be identified with the arrow of time what means that the vector field is future-oriented.
%Since we treat an observer as the vector field that vector field can span a one-dimensional timelike distrubution. It leads
%to 3+1 decomposition of the spacetime manifold where the three-dimensional distribution is spacelike and sometimes called a hypersurface.

Before proceeding further, let us answer the question of when the one-dimensional foliation spanned by $u$ is totally geodesic. This can
be easily done by studying the auto-parallel (geodesic) equation
\begin{equation}\label{auto}
    u^\beta u_{\alpha ;\beta}=0\,,
\end{equation}
where $u_{\alpha ;\beta}=\nabla_\beta u_\alpha$ denotes the Levi-Civita covariant derivative of $u$. Thus introducing the
acceleration vector $\dot u_\alpha = u^\beta u_{\alpha ;\beta}$ one can conclude that the vanishing of $\dot u^\alpha$ is equivalent
to the geodesic equation (\ref{auto}). One should notice that $\dot u_\alpha$ is, in fact, a three-vector,
since $u^\alpha \dot u_\alpha=0$.\\
 In general, one can decompose the space components of the two-tensor \(u_{\alpha;\sigma}\) into irreducible parts with respect to the
three-dimensional orthogonal group:
\begin{equation}
 \tilde u_{\alpha;\beta}=h^\sigma_{\beta}u_{\alpha;\sigma}=\omega_{\alpha\beta}+\sigma_{\alpha\beta}+\frac{1}{3}\Theta h_{\alpha\beta},
\end{equation}
where \(\omega_{\alpha\beta}\) denotes its antisymmetric part, \(\sigma_{\alpha\beta}\) is the traceless symmetric component and finally
  \(\Theta\) stands for the trace. This is a kinematical decomposition
\footnote{The dynamical equation is known as Raychaudhuri equation (see e.g. \cite{poisson,slobodeanu}).}. Using (\ref{proj}) we shall obtain \cite{ehlers,eckart}:
\begin{equation}
 u_{\alpha;\beta}=\omega_{\alpha\beta}+\sigma_{\alpha\beta}+\frac{1}{3}\Theta h_{\alpha\beta}-\dot{u}_\alpha u_\beta.
\end{equation}
There is a well-known interpretation of the observer in terms of relativistic hydrodynamics, treating it as a flow of material
points constituting a (perfect) fluid (continuous medium), with the world lines being particle trajectories: one line of the flow passes through
every point \(x^\alpha\) of a certain spacelike (possibly bounded) region in spacetime. Accordingly, the tensor \(u_{\alpha;\sigma}\)
determines the rate of change in the position of one point with respect to the other one in the material \cite{pleb}.\\
Keeping in mind this fluid analogy, each irreducible component of the projected tensor $u_{\alpha ; \beta}$ admits a physical
interpretation which is contained in self-explanatory and intuitive names (for more detailed explanations see e.g. \cite{ehlers,dem,pleb}).
In a more explicit form, one has to take into account the following three-dimensional quantities:
\begin{align}\label{ehl}
 \omega_{\alpha\beta}&=u_{[\alpha;\beta]}+\dot{u}_{[\alpha}u_{\beta]}&\text{is the rotation tensor,}\\
 \sigma_{\alpha\beta}&=u_{(\alpha;\beta)}+\dot{u}_{(\alpha}u_{\beta)}-\frac{1}{3}\Theta h_{\alpha\beta}&\text{is the shear tensor,}\\
 \Theta&=u^\alpha_{\,;\alpha}&\text{is the expansion scalar,}\\
 \dot{u}^\alpha&=u^\beta u^\alpha_{\,;\beta}&\text{is the acceleration vector.}
\end{align}
It is more convenient to use the scalars
\begin{equation}
 \dot{u}\equiv(\dot{u}_\alpha\dot{u}^\alpha)^{\frac{1}{2}},\;\;\;\;\omega\equiv(\frac{1}{2}\omega_{\alpha\beta}\omega^{\alpha\beta})^{\frac{1}{2}},\;\;\;\;
 \sigma\equiv(\frac{1}{2}\sigma_{\alpha\beta}\sigma^{\alpha\beta})^{\frac{1}{2}}.
\end{equation}
These are non-negative and vanish at the same time as their corresponding tensors. An observer is rotation-free, shear-free or
expansion-free when \(\omega=0\), \(\sigma=0\) or \(\Theta=0\) respectively.
If all quantities vanish, then the observer is called rigid.\\
It is worth mentioning that the observer (four-velocity) field $u^\alpha$ can be used to construct the energy momentum tensor of an
ideal (incompressible) fluid,
\begin{equation}\label{if}
T_{\alpha\beta}= (p+\rho)\,u_{\alpha}u_{\beta}+ p g_{\alpha\beta}\,,
\end{equation}
where the matter density  $\rho$ and the pressure $p$ are internal fluid parameters determining its thermodynamical behavior. The same
energy momentum tensor
treated on the right-hand side of Einstein's equations as the source of the gravitational field influences the metric.
This suggests possible relationships between metric and  fluid  observer, which are an interesting subject for future
research (see e.g. \cite{giulini,slobodeanu}).\\
Now we are ready to classify all almost-product structures related to relativistic observers in gravitational spacetimes. As we have already
mentioned, the tensor \(h^\alpha_{\,\beta}\) projects on a three-dimensional subspace while
\(-u^\alpha u_\beta=\delta^\alpha_\beta -h^\alpha_\beta\) projects on the one-dimensional complementary distribution spanned by \(u^\alpha\).
It turns out that the difference:
\begin{equation}\label{struc}
 P^\alpha_{\beta}=h^\alpha_\beta-(-u^\alpha u^\beta)=\delta^\alpha_{\beta}+2u^\alpha u_\beta.
\end{equation}
represents an almost-product structure compatible with the metric \(g\) \footnote{It is easy to see that \(P\) satisfies the conditions
\(P^2=I\), as well as (\ref{comp}).}. Now the almost-product structure (\ref{struc}) can be used
 to encode the observer \(u^\alpha\).\\
Since the issue of one-dimensional distributions have already been solved, we should concentrate on the three-dimensional one.
One has 4 conditions to be imposed on $P$ (see Definition 1 and Theorem 2). The umbilical case \(D_3\), after some manipulations, produces
\begin{equation}
 u_{(\alpha;\beta)}+\dot{u}_{(\alpha}u_{\beta)}=\frac{1}{3}\sum^3_{i=1}e^\beta_i u_{\alpha;\beta}e^\alpha_i.
\end{equation}
Here \(\{e_i\}_{i=1}^3\) denotes a local orthonormal frame of \(D\). We notice that the sum of the right
hand side is a three-dimensional trace of the tensor \(u_{\alpha;\beta}\), thus \(D_3\) is equivalent to the vanishing of the shear tensor.\\
The condition \(D_2\) (a minimal distribution) leads to
\begin{equation}
 \sum^3_{i=1}e^\beta_i u_{\alpha;\beta} e^\alpha_i=0\,,
\end{equation}
which is equivalent to the vanishing a scalar of expansion.\\
For a geodesic distribution (\(D_1\)), one obtains
\begin{equation}\label{killing}
 u_{(\alpha;\beta)}+\dot{u}_{(\alpha}u_{\beta)}=0\,,
\end{equation}
which is equivalent to the vanishing of both characteristics: shear and expansion. In the free falling case $(\dot{u}=0)$ the condition (\ref{killing}) denotes that the normalized timelike vector $u^\alpha$ is a Killing vector for the metric $g_{\mu\nu}$.\\
Similarly, one can show that the integrability condition (\(F\)) reduces to
\begin{equation}
  u_{[\alpha;\beta]}+\dot{u}_{[\alpha}u_{\beta]}=0\,,
\end{equation}
which means vanishing of a rotation.\\
The final results are presented in the Table $1$ and Table $2$. The first one concerns accelerated \(\dot{u}\neq0\) observers with all possibilities
for the three-dimensional distribution taken into account. Similarly, the second table concerns free-falling
observers (\(\dot{u}=0\)). The bracket \((1,3)\) in the first column indicates that
the first symbol is for the one-dimensional distribution and
the second for the three-dimensional one. The tables also contain the physical interpretations for each class of almost-product
Lorentzian manifolds.

\begin{table}[p]
\caption{Accelerated observers}
%        \centering
%        \rotatebox{90}{
%                \begin{minipage}{\textheight}
%\begin{center}
    \begin{tabular} {| >{\centering}p{0.9cm} | c | >{\centering}p{2.39cm} | >{\centering}p{2.4cm}|}
    \hline
    \textbf{Class\newline (1,3)}&  \raisebox{-.5\height}{Accelerated observers \(\dot{u}\neq0\)} & \textbf{Physical meaning} & \raisebox{-.5\height}{\textbf{3-distribution}} \tabularnewline \hline
    \raisebox{-.5\height}{\((F,-)\)} & \raisebox{-.5\height}{\(u_{\alpha;\beta}=\sigma_{\alpha\beta}+\omega_{\alpha\beta}+
 \frac{1}{3}\Theta h_{\alpha\beta}-\dot{u}_\alpha u_\beta\)}  & \raisebox{-.5\height}{------}  & non-integrable \textbf{distribution}
    \tabularnewline \hline
    \((F,D_2)\) & \(\Theta=0\;\Rightarrow\;%\sum^3_{i=1}e^\alpha_i(u^\beta\,_{;\alpha}u_\gamma+u^\beta
   % u_{\gamma;\alpha})e^\gamma_i=0\)
u_{\alpha;\beta}=\sigma_{\alpha\beta}+\omega_{\alpha\beta}
 -\dot{u}_\alpha u_\beta\)
& expansion-free & minimal\tabularnewline \hline
\((F,D_3)\) & \(\sigma=0\;\Rightarrow\;\)%\;\;X^\beta Y^\gamma(u_{(\beta;\gamma)}+\dot{u}_{(\beta}u_{\gamma)})
 %=\frac{4}{3}X^\mu Y_\mu\sum^3_{i=1}e^\beta_i(u_{\gamma;\beta})e^\gamma_i\)
\(u_{\alpha;\beta}=\omega_{\alpha\beta}+
 \frac{1}{3}\Theta h_{\alpha\beta}-\dot{u}_\alpha u_\beta\)
 & shear-free & umbilical\tabularnewline \hline
  \raisebox{-.5\height}{\((F,D_1)\)} & \raisebox{-.5\height}{\(\Theta=\sigma=0\;\Rightarrow\;\)
    \(u_{\alpha;\beta}=\omega_{\alpha\beta}-\dot{u}_\alpha u_\beta\)}  & shear-free \& expansion-free & \raisebox{-.5\height}{geodesic}
    \tabularnewline
    \hline\hline
\((F,F)\) & \(\omega=0\;\Rightarrow\;u_{\alpha;\beta}=\sigma_{\alpha\beta}+
 \frac{1}{3}\Theta h_{\alpha\beta}-\dot{u}_\alpha u_\beta\) &  rotation-free & \textbf{foliation} \tabularnewline \hline
\raisebox{-.5\height}{\((F,F_2)\)} &
\raisebox{-.5\height}{\(\omega=\Theta=0\;\Rightarrow\;u_{\alpha;\beta}=\sigma_{\alpha\beta}
 -\dot{u}_\alpha u_\beta\)} & rotation-free \& expansion-free & \raisebox{-.5\height}{minimal}\tabularnewline \hline
\raisebox{-.5\height}{\((F,F_3)\)} & \raisebox{-.5\height}{\(\omega=\sigma=0\;\Rightarrow\;
u_{\alpha;\beta}=\frac{1}{3}\Theta h_{\alpha\beta}-\dot{u}_\alpha u_\beta\)}
& rotation-free \& shear-free &  totally umbilical \tabularnewline \hline
\raisebox{-.5\height}{\((F,F_1)\)} & \raisebox{-.5\height}{\(u_{\alpha;\beta}=-\dot{u}_\alpha u_\beta\)} &  \raisebox{-.5\height}{rigid}
 & totally geodesic \tabularnewline \hline
\end{tabular}
%\end{center}
%\end{minipage}
%        }
\end{table}
\begin{table}[p]
\caption{Free-falling observers}
%        \centering
%        \rotatebox{90}{
%                \begin{minipage}{\textheight}
%\begin{center}
\begin{tabular}{ | >{\centering}p{1.10cm} | c | >{\centering}p{2.50cm} | >{\centering}p{2.45cm} |}
\hline
\textbf{Class\newline (1,3)}& \raisebox{-.5\height}{Geodesic (free falling) observers \(\dot{u}=0\)} &  \textbf{Physical meaning} & \raisebox{-.5\height}{\textbf{3-distribution}} \tabularnewline \hline
\raisebox{-.5\height}{\((F_1,-)\)} & \raisebox{-.5\height}{\(u_{\alpha;\beta}=u_{[\alpha;\beta]}+\sigma_{\alpha\beta}+
 \frac{1}{3}\Theta h_{\alpha\beta}\)} &  \raisebox{-.5\height}{geodesic} & non-integrable \textbf{distribution} \tabularnewline \hline
\raisebox{-.5\height}{\((F_1,D_2)\)} & \raisebox{-.5\height}{\(\Theta=0\;\Rightarrow\;u_{\alpha;\beta}=u_{[\alpha;\beta]}+\sigma_{\alpha\beta}\)}
 & geodesic expansion-free & \raisebox{-.5\height}{minimal} \tabularnewline \hline
\raisebox{-.5\height}{\((F_1,D_3)\)} &  \raisebox{-.5\height}{\(\sigma=0\;\Rightarrow\;u_{\alpha;\beta}=u_{[\alpha;\beta]}+
 \frac{1}{3}\Theta h_{\alpha\beta}\)} &  geodesic shear-free & \raisebox{-.5\height}{umbilical} \tabularnewline \hline
\raisebox{-1.5\height}{\((F_1,D_1)\)} & \raisebox{-1.5\height}{\(\sigma=\Theta=0\;\Rightarrow\;u_{\alpha;\beta}=u_{[\alpha;\beta]}\)} & geodesic shear-free \& expansion-free & \raisebox{-1.5\height}{geodesic} \tabularnewline \hline\hline
\raisebox{-.5\height}{\((F_1,F)\)} & \raisebox{-.5\height}{\(\omega=0\;\Rightarrow\;u_{\alpha;\beta}=\sigma_{\alpha\beta}+
 \frac{1}{3}\Theta h_{\alpha\beta}\)} & geodesic rotation-free & \raisebox{-.5\height}{\textbf{foliation}} \tabularnewline \hline
\raisebox{-1.5\height}{\((F_1,F_2)\)} &  \raisebox{-1.5\height}{\(\omega=\Theta=0,\;\Rightarrow\;u_{\alpha;\beta}=\sigma_{\alpha\beta}\)}
 & geodesic rotation-free \& expansion-free & \raisebox{-1.5\height}{minimal} \tabularnewline \hline
\raisebox{-1.5\height}{\((F_1,F_3)\)} &  \raisebox{-1.5\height}{\(\omega=\sigma=0\;\Rightarrow\;u_{\alpha;\beta}= \frac{1}{3}\Theta h_{\alpha\beta}\)}
 & geodesic rotation-free \& shear-free & \raisebox{-1.5\height}{totally umbilical} \tabularnewline \hline
\raisebox{-.1\height}{\((F_1,F_1)\)} &  \raisebox{-.1\height}{\(u_{\alpha;\beta}=0\)} & geodesic rigid & \raisebox{-.1\height}{totally geodesic} \tabularnewline \hline
\end{tabular}
%\end{center}
%\end{minipage}
%        }
\end{table}

\section{Illustrative examples}
\subsection{Minkowski spacetime}
The most extreme case in Naveira's classification is \((F_1,F_1)\) class, i.e. both distributions are totally geodesic foliations.
In Minkowski spacetime the metric is flat in the Cartesian coordinate system (\(\eta_{\mu\nu}=\text{diag}(-1,1,1,1)\)), so one can replace the covariant derivatives with partial ones.
Then the \((F_1,F_1)\) case becomes just \(u_{\alpha,\beta}=0\). There exists a solution in the form of a constant vector. In fact, any constant
timelike vector field can be changed by a linear transformation of coordinates (i.e. Lorentz transformation) into
\begin{equation}\label{mink}
 u^\alpha=\left[1,0,0,0\right].
\end{equation}
Such a vector field is a canonical inertial observer in Minkowski spacetime. A less restrictive class is \((F,F_1)\), which implies that the observer should
accelerate. An example of such an observer is Rindler's:
\begin{equation}
 u^\alpha=\left[\frac{x}{x^2-t^2},\frac{t}{x^2-t^2},0,0\right],
\end{equation}
for whom only a part of Minkowski space is available. The only non-vanishing characteristic is the acceleration
 \begin{equation}
\dot{u}=(x^2-t^2)^{-1/2},
\end{equation}
which is constant along each trajectory. Again, by introducing adapted (Rindler's) coordinates one can simplify expressions.\\
Let us consider the rotating observer in the $(x,y)$ plane,
\begin{equation}\label{rotating}
 u^\alpha=\left[\sqrt{2},\frac{-y}{\sqrt{x^2+y^2}},\frac{x}{\sqrt{x^2+y^2}},0\right]\,,
\end{equation}
belonging to the class \((F,D_2)\) with the following characteristics:
\begin{align}
\Theta&=0,\\
 \dot{u}&=(x^2+y^2)^{-1/2},\\
\omega&=\frac{\sqrt{2}}{2}(x^2+y^2)^{-1/2},\\
\sigma&=\frac{\sqrt{2}}{2}(x^2+y^2)^{-1/2},
\end{align}
constant along particles trajectories.
The last example for Minkowski spacetime (in the spherical coordinates \(g_{\mu\nu}=\text{diag}(-1,1,r^2,r^2 \sin^2{\theta})\)) is the
observer
\begin{equation}\label{minksfer}
 u^\alpha=\left[\frac{r}{\sqrt{r^2-t^2}},\frac{t}{\sqrt{r^2-t^2}},0,0\right].
\end{equation}
It turns out that observer (\ref{minksfer}) has the following characteristics:
\begin{align}
 \dot{u}&=(r^2-t^2)^{-1/2},\\
\omega&=0,\\
\Theta&=\frac{2t}{r}(r^2-t^2)^{-1/2},\\
\sigma&=\frac{\sqrt{3}t}{3r}(r^2-t^2)^{-1/2},
\end{align}
and belongs to the class \((F,F)\).

\subsection{Schwarzschild spacetime}
There is no observer belonging to the class \((F_1,F_1)\) in Schwarzschild spacetime. It should satisfy
the sixteen equations \(u_{\alpha;\beta}=0\), which turn out to be inconsistent.\\
Let us consider the observer
\begin{equation}\label{stat}
 u^\alpha=\left[(1-\frac{2M}{r})^{-\frac{1}{2}},0,0,0\right].
\end{equation}
The only non-vanishing characteristic is the acceleration \(\dot{u}=\frac{M}{r^2}(1-\frac{2M}{r})^{-1/2}\),
which implies that observer (\ref{stat}) belongs to the class \((F,F_1)\).\newline
The geodesic observer \cite{poisson} is of the form
\begin{equation}
 u^\alpha=\left[(1-\frac{3M}{r})^{-1/2},0,\sqrt{\frac{M}{r^2(r-3M)}},0\right].
\end{equation}
It shows a singular expansion at the north and south poles:
\begin{equation}
 \Theta=\sqrt{\frac{M}{r^2(r-3M)}}\cot\theta.
\end{equation}
The rotation and shear scalars are
\begin{align}
\omega&=\frac{1}{4}\sqrt{\frac{M}{r^3}}\left(\frac{1-6M/r}{1-3M/r}\right),\\
\sigma&=\sqrt{\frac{27M\sin^2{\theta(r-2M)^2}+16Mr(r-3M)\cos^2{\theta}}{48r^3(r-3M)^2\sin^2{\theta}}}.
\end{align}
It belongs to the $(F,-)$ class, so for this observer, as well as for (\ref{rotating}) of Minkowski spacetime, there is no
three-dimensional orthogonal distribution
providing a foliation of the spacetime manifold $(M,g)$ i.e. there are no three-dimensional equal-time subspaces relative to these
observers.

\end{document}